\def\year{2019}\relax
\documentclass[letterpaper]{article} 
\usepackage{aaai19}  
\usepackage{times}  
\usepackage{helvet} 
\usepackage{courier}  
\usepackage[hyphens]{url}  
\usepackage{graphicx} 
\usepackage{graphicx}  
\usepackage{cleveref}

\usepackage{xspace}
\def\BSDS{{\sc bs:ds}\xspace}

\title{Data Science as a Route to AI for Middle- and High-School Students}
\author{
  Shriram Krishnamurthi\textsuperscript{\rm 1},
  Emmanuel Schanzer\textsuperscript{\rm 1},
  Joe Gibbs Politz\textsuperscript{\rm 2}, \\
  \Large \textbf{
    Benjamin S. Lerner\textsuperscript{\rm 3},
    Kathi Fisler\textsuperscript{\rm 1},
    Sam Dooman\textsuperscript{\rm 4}}
  \\
  \textsuperscript{\rm 1} Brown University and Bootstrap \\
  \textsuperscript{\rm 2} UCSD and Bootstrap \\
  \textsuperscript{\rm 3} Northeastern University and Bootstrap \\
  \textsuperscript{\rm 4} Work done at Brown University; current affiliation: Down Dog Yoga \\
  Contact email: schanzer@bootstrapworld.org
}

\begin{document}

\maketitle

\begin{abstract}
  The Bootstrap Project's Data Science curriculum has trained
  about 100 teachers who are using it around the country. It is
  specifically designed to aid adoption at a wide range of
  institutions. It emphasizes valuable curricular goals by drawing on
  both the education literature and on prior
  experience with other computing outreach projects. It embraces
  ``three P's'' of data-oriented thinking: the promise, pitfalls,
  and perils. This paper briefly describes the curriculum's
  design, content, and outcomes, and explains its value on the
  road to AI curricula.
\end{abstract}

\noindent
Modern AI is heavily data-centric. Becoming a successful student of AI
requires a reasonable facility with data in several dimensions:
writing code to process it; employing statistics to summarize and
understand it; and engaging with society to understand both its power
and its potential for harm.

In this paper we outline the Bootstrap:Data Science (\BSDS) curriculum.
This curriculum is part of the Bootstrap
project (\url{https://www.bootstrapworld.org/}), which
is over a dozen years old and is one of the leading curriculum
providers for middle- and high-school computing in the
USA.\footnote{The term ``Bootstrap'' in this paper will always refer
  to this \emph{project}, and never refers to the statistical or
  computing \emph{concepts} called bootstrapping.} Bootstrap has been
particularly successful in its oldest curriculum, which teaches key
algebra concepts~\cite{sfk:bsa-scaff-unscaff-wp}, through the strategy of \emph{integration}:
rather than teaching computing directly, it
piggybacks atop existing courses and expertise in
schools. \BSDS leverages this visibility in school systems and
awareness amongst both computing and math teachers.

\section{AI and Data Science}

\BSDS is primarily a \emph{data science} curriculum, not an AI
curriculum. Nevertheless, we believe it is deeply connected to the
subject of this workshop.

First, many of the central concepts---writing programs over data,
using these programs to answer questions, and using statistical
techniques for this task---are common to both. Second, while \BSDS
does not currently handle all the forms of data found in AI---such as
plain text or streaming data---it provides a strong foundation atop
which these materials can be added (possibly with collaborators who
wish to work with us). Finally, because a facility with data is a
prerequisite for many AI systems, \BSDS can serve as a prerequisite
for AI courses that educators wish to build.

Furthermore, bringing AI to schools requires far more than just
``teaching AI''. There are numerous logistical, pedagogic,
developmental, and cognitive issues that need to be addressed. For the
foreseeable future, most US schools do not have room to add whole AI
classes. Students need to feel confident tackling this material,
rather than opting out of it as being not for ``people like'' them (as
reinforced by both leaders and media portrayals). They need to gain
comfort with the techniques, which can be demanding, and depend on a
wide range of STEM skills (which are already often in short
supply). Finally, many of these issues also apply at the level of
teachers, for their preparation and support.

Because \BSDS already addresses many of these issues, it enables educators
to focus more on their core ideas and less on these logistics, which would
otherwise also be their bailiwick. Indeed, imagine if an AI course
could just \emph{assume} every incoming student already knew basic
programming, statistics, and data analysis---think how far it could go!
We therefore hope this paper will
spur interesting conversations and potentially also collaborations
with AI researchers.

\section{Curricular Goals and Constraints}
\label{s:goals}

As outlined in a recent CACM paper~\cite{skf:curriculum-succeed}, the Bootstrap project has
three goals across all curricula. These bear repeating here, because
they significantly impact the design and content of \BSDS:
\begin{description}

\item[Equity] Equity is the goal that the curriculum be accessible
  across various student communities. There are numerous
  aspects to equity that \BSDS pays attention to; here are a few of
  them:
  \begin{itemize}

    \item Specialized offerings like opt-in courses, after-school
      courses, etc.\ have self-selecting student groups: either those
      advantaged to take it (e.g., those who don't need to hold down
      an after-school job) or those who identify with the course
      (e.g., the gender and ethnicities already present in such
      courses, which are highly skewed). Thus, it is best to try to
      put material into courses that all students in a school already
      take; required math courses have been a particularly productive
      avenue for Bootstrap. These have the same gender and
      underrepresentation ratios as the school itself.

    \item Putting material in required courses means addressing the
      learning needs of \emph{all} students in that course. As one
      illustrative example: though a
      notable percentage of students are visually impaired (\url{https://nfb.org/resources/blindness-statistics}),
      many computing courses are largely or entirely
      inaccessible to them. For this reason, Bootstrap has spent
      significant effort making both the IDEs and the programming
      languages accessible to blind and other visually impaired
      students~\cite{sbk:accessible-ast-blocks}.

    \item Just because material is introduced into a course does not
      mean it is \emph{appealing} to all the students in the
      course. Giving students the ability to customize their learning,
      without disrupting the material's learning goals, is an
      important way to make it appeal to broad populations. Our prior
      study in Bootstrap:Algebra~\cite{skf:creativity-bootstrap} shows that even a \emph{small}
      amount of carefully targeted customization is enough to make
      students feel a great deal of ownership and pride in their
      work. Though we have not yet formally studied the same issue in
      \BSDS, we use the same ``formula'', as described later in this paper.

  \end{itemize}

\item[Rigor] Rigor should be fairly self-evident, but is often missing
  in many curricula that aim for equity. There is an undertone in some
  middle- and high-school computing curricula that students reject
  rigorous material, and the frustrations of programming (not the
  same); that including these will therefore make students less
  interested in the subject; and that these should therefore be
  minimized. In contrast, \BSDS embraces the concept of productive
  struggle~\cite{hg:eff-class-math-stud-learn}, and uses program design methods~\cite{fffk:htdp} that employ
  Bruner's notion of scaffolding~\cite{wbr:role-tutoring-prob-solv} and Vygotsky's concept of
  zones of proximal development~\cite{lsv:mind-in-society} to help a student make steady
  and measurable progress without compromising the content.

\item[Scale] Finally, the Bootstrap project cares deeply about
  scale. It is relatively easy to create highly rigorous curricula in
  very specialized settings: the experience of several of this paper's
  authors, who teach introductory computing at their universities, is
  that a small number of schools have (almost surprisingly) deep
  computing content covering several topics, especially AI.

  However, these approaches do not currently scale: they depend on
  hardware (that may be expensive but even more importantly needs to
  be \emph{maintained}), trained teachers (when there is already a
  significant national shortage of computing teachers), time in the
  curriculum and space in the school, etc. In fact, at many public
  schools in the US (the primary audience of Bootstrap), even more
  basic assumptions cannot be met: many schools have locked-down
  computers (to prevent viruses), and hence cannot install new
  software; and at many schools, students cannot be assured daily
  computer access. While some situations may be nearly impossible to
  address, we feel that scaling (by definition) means tackling the
  needs of roughly the 20th percentile, not the 90th.

\end{description}

Success also requires attention to the factors that influence teacher
uptake. These include teacher preparation, alignment with standards,
making the material engaging to different groups of students, respect
for the amount of time teachers can make in a class, and so on.

\section{Curriculum Description}

\subsection{Major Components}

At a high level, the \BSDS curriculum has three main aspects:
\begin{description}

\item[Introduction to Programming] A central part of our curriculum is
  to program over data. For now, \BSDS focuses on \emph{structured}
  data.\footnote{We very much wish to incorporate various forms of
    unstructured data (such as raw text) in the future, but we believe
    this requires more attention paid to the statistics components.}
  We focus on tabular data, which are a useful abstraction over many
  data sources, from streams of sensor readings to census bureau
  reports and much more. We primarily use Pyret (\url{https://www.pyret.org/}), which can be
  thought of as a simplified, more student-friendly version of Python.
  It has a sophisticated~\cite{bnpkg:stopify} Web implementation that hides
  several complexities of the Web platform from students.

  Students entering \BSDS increasingly have \emph{some} programming
  familiarity, such as using Scratch~\cite{Resnick:2009:SP:1592761.1592779} in primary school. Some
  students, of course, have not had this prior exposure. Even for
  those who have, however, in our experience, these exposures tend to
  be quite superficial, do not transfer particularly well to textual
  programming (see~\cite{gpc:design-blended}), and are highly
  data-impoverished, leaving students with no real facility for
  thinking about processing data. Thus, we effectively have to start
  programming all over again, which we do not find problematic since
  this is easier than trying to cater to a variety of backgrounds.

\item[Introduction to Statistics] \BSDS is used in several subjects:
  general math, statistics, business, and social studies. In
  many of these, it is necessary to also include introductory material
  on statistical reasoning. Over time, we have found this material is
  needed not only for the students but also to train some of the
  teachers who use the curriculum. While we can assume a basic
  knowledge of the mean, median, and mode, even regression is not a
  topic we can assume. In fact, the more statistics material we
  include rather than assume, the more demand we find for it from
  teachers, and hence have to keep expanding it.

\item[The Three P's] As noted earlier, our curriculum revolves around
  ``three P's'': promise, pitfalls, and perils. We discuss each below:

  \begin{description}

  \item[Promise] The promise of data-oriented methods are
    not at all universally familiar to teachers and
    students in schools. Thus, a major portion of the curriculum is
    having students wrestle with real-world data sets and learn things
    from them. Both teachers and students find it revelatory that they
    have the power to formulate hypotheses and test them against data.

  \item[Pitfalls] Programming with large quantities of data also
    introduces pitfalls. There is a well-known tendency to
    assume an output is correct because it came from a
    computer. Furthermore, so long as an output does not wildly
    deviate from the programmer's expectations (which may themselves
    be highly, and even implicitly, biased), there is not much reason
    to question it. Instead, \BSDS asks teachers and students to
    create representative data samples and formulate tests on these,
    ensuring that their program is producing the desired output on
    \emph{known} data sets before trusting their output on unknown
    ones. 

  \item[Perils] Over the past several years, journalists of various
    stripes have done a great deal to expose some of the potential
    perils of data-oriented thinking (e.g., the seminal ProPublica
    study on parole~\cite{almk:machine-bias-crime}). As data-driven AI systems are embedded
    even deeper into society (e.g., China's social credit system),
    it is vital for students to understand the role of these
    systems in their lives and the problems they can cause. For many
    \BSDS students, these issues are deeply personal, as they may
    even affect immediate family members. \emph{We believe every
      data-oriented curriculum has a moral obligation to teach not
      only the methods but also the responsibilities that come with
      data-centricity.}

  \end{description}

\end{description}
In addition, the curriculum has two important facets that have proven
invaluable:
\begin{description}

\item[Notice and Wonder] Real data demand reflection:
  to understand its structure, meaning, value, and problems. However,
  simply telling students to ``think'' about data is not likely to be
  effective. We instead structure this thinking through the
  technique of ``notice and wonder''~\cite{mr:powerful-problem-solving}, which carries
  over superbly to reasoning about data. This requires asking students
  to study the data to first see what they notice; and then using
  their observations to ask what they wonder. Having a class
  collectively share their observations and questions is a revealing
  experience, and gives students different ways to shine: a student
  may not be very facile at programming, but may be particularly good
  at observing weaknesses in a data set.

\item[Authenticity] Finally, \BSDS is attractive to teachers because
  we offer \emph{authentic} embeddings into
  different contexts. For instance, as noted above, some
  of our teachers teach social studies. Social studies is a superb area for
  introducing data-oriented thinking. However, most of these
  teachers do not much care about
  \emph{programs}; their unit of currency is the \emph{research
    report}. We present programs (and statistics) as \emph{yet another
    way of answering questions}, a concept both familiar and welcome
  to these teachers. Programs then become just one more tool in their
  arsenal, which already includes concepts like literature reviews and
  surveys. The curriculum thus leads to the creation of a report:
  starting with a question, then finding relevant data, then
  determining the statistics and programming them\dots but then also
  closing the circle, returning to answer the question and justifying the
  answer based on the computed statistics. This has the authenticity
  that is necessary for teachers to comfortably incorporate seemingly
  alien material into their classes, thus greatly increasing the reach
  of \BSDS.

\end{description}

\subsection{Module Structure}

\begin{figure}[t]

\begin{tabular}{c|p{1.3in}|p{1.3in}}

  \hline
  
  {\bf Unit}
  & {\bf Data}
  & {\bf Programming} \\

  \hline
  
  1
  &
    Introduction to data: tables, categorical, and quantitative
    data.
  &
    Introduction to data types in programming. \\

  2
  &
    Learning to ask questions about data.
  &
    Defining functions, filtering. \\

  3
  &
    Preparing logical subsets of data. Focus on categorical data.
  &
    Defining and using filter functions. Displaying data. \\

  4
  &
    Histograms and bar-charts (and their
    difference); manual construction; interpretation.
  &
    Doing these programmatically. \\

  5
  &
    Central tendencies and spread.
  &
    Doing these programmatically. \\

  6
  &
    Table manipulation. Trusting data.
  &
    Methods for building and transforming tables. \\

  7
  &
    Scatter plots. Correlations.
  &
    Doing these programmatically. \\

  8
  &
    Linear regression. More on correlations.
  &
    Doing these programmatically. \\

  9
  &
    Threats to validity.
  &
    Wrapping up projects and reports. \\

  \hline
  
\end{tabular}

\caption{\BSDS Modules}
\label{f:modules}

\end{figure}

The curriculum currently has 9 units
(\cref{f:modules}). Each unit is divided into data-oriented and
programming-oriented learning, which proceed roughly in lock-step.

A key idea is personalization: early in the curriculum, students
choose a data set of interest to them. \emph{We believe that all
  students are inherently data scientists}: even a student who ``hates
math'' will passionately scour for and deploy statistics to score a
point about, say, their favorite sports player. The key is to find the
data sets that students are passionate about. Thus, a great deal of
effort in \BSDS goes towards identifying and curating suitable data
sets, from income and schools to cancer rates to movies to sodas and
cereals to Pok\'emon characters to sports.

One topic that is \emph{not} currently a part of \BSDS is data
cleansing. We introduce the idea and have students understand the
messy nature of real-world data, but the set of both programming and
statistical techniques needed to effectively cleanse data are well
beyond middle- and most early high-school courses. Instead, we provide
a set of curated data sets that have already been cleansed by us, and
work with teachers to add new data sets of interest to them.

\section{Deployment}
\label{s:deployment}

\BSDS has trained about 100 teachers so far, several of whom are in
various stages of deployment. The curriculum is currently in use in
six different US states and has some international interest as
well. The entire material is available for free
from \url{https://www.bootstrapworld.org/materials/data-science/}.

\section{The Role of Standards}

Many teachers need to align their classroom content with
standards. Currently, these standards say little to nothing about
AI. However,
the CSTA standards (\url{https://www.csteachers.org/page/standards})
and K-12 Framework (\url{https://k12cs.org})
highlight data and analysis as a major
content theme. Required elements within this theme include presenting
and making claims from data, cleaning data, and creating and refining
models based on data. Some of this content is highly relevant to learning
about AI.  However, the \emph{practices} side of the standards---the part that
deals with the habits of thinking and behavior that underlie effective
work in computing---fall short in ways that a good data
science curriculum should mitigate. For instance, designing (and using!)\ test data,
and making observations about data prior to drawing conclusions from it,
are key skills that underlie responsible AI. The \BSDS curriculum
fills these gaps, and other curricula should endeavor to as well.

\section{A Caution}

Collegiate educators are used to AI being in the middle of a large DAG of
courses. However, this is unlikely to happen at most schools.
Their AI course could 
become the \emph{only} ``computer science'' course
in many schools. As the primary or even sole reference
point, the habits it encourages are the ones that students will
carry forward.

This means an ends-at-all-costs approach will have bad
side-effects. For instance, the rough-and-ready software style
espoused in some projects---and not prohibited by
standards practices---will then become students'
default mode of operation. \BSDS recognizes this danger and therefore
builds topics like systematic software construction, paying attention
to validation, etc., into the materials. We exhort other curriculum
designers to attend to these same issues.

\section{Ongoing and Future Work}

The current module structure of \BSDS is largely an artifact of our
pathway to adoption. Presenting whole-semester, all-or-nothing options
to teachers is usually a non-starter. Instead, the
Bootstrap curricula begin as a collection of modules that
teachers can intersperse through their existing courses while
meeting state and national standards.

As the curricula grow in adoption, however, we have the freedom to
take up more time. For instance, some regions and states are now
looking at using Bootstrap curricula over
a year or even two years across math classes. In these settings,
our hope is to expand \BSDS on three fronts: statistics, threats to
validity, and improved testing and validation. Of course, some
teachers prefer to go into more advanced programming as well: for
instance, methods for tabular operations are effectively higher-order
functions, a topic that a few teachers want to explore in its own
right.

A separate direction, which is unlikely to directly impact AI but will
likely leave students much better prepared for it, is new 
material we are piloting for middle-school history.
Students who have performed
programming-driven data analysis several times over the course of
their history curriculum will likely come into a future AI-centric
class far better prepared to start at a more advanced stage and to
take on its challenges.

\section{ Acknowledgments}

We thank Leigh Ann DeLyser for her support for \BSDS over several
years. We are grateful for support from the US National
Science Foundation and from Bloomberg, the Robin Hood Foundation,
and the Seigel Family Foundation.

\small

\bibliography{shriram,extra}
\bibliographystyle{aaai}

\end{document}